\documentclass[12pt,twoside]{article}
\usepackage{fleqn,espcrc1}

\usepackage{graphicx}

\def\thebib#1{
 \list
 {\arabic{enumi}}{\settowidth\labelwidth{[#1]}\leftmargin\labelwidth
 \advance
 \leftmargin\labelsep
 \setlength{\parsep}{0mm}%
 \setlength{\itemsep}{0mm}%
\usecounter{enumi}}
 \def\newblock{\hskip .11em plus .33em minus .07em}
 \sloppy\clubpenalty4000\widowpenalty4000
 \sfcode`\.=1000\relax}

\def \invisible{\mbox{$\rule{0mm}{1mm}$}}
\def \mathbox(#1){\invisible\ifmmode{{#1}}\else{\mbox{${#1}$}}\fi}
\def \mbf(#1){\mbox{\boldmath{$#1$}}}

\def\III{{\rm I\thinspace I\thinspace I}}

\newcommand{\half}{\mbox{$\frac{1}{2}$}}

\newcommand{\psibar}{\bar{\psi}}

\def \invisible{\mbox{$\rule{0mm}{1mm}$}}
\def \mathbox(#1){\invisible\ifmmode{{#1}}\else{\mbox{${#1}$}}\fi}
\def \mbf(#1){\mbox{\boldmath{$#1$}}}

\def\rmN{{\rm N}}

\def\SLJ(#1,#2,#3){\mathbox({^{#1}\!#2_{#3}})}

\def\FRAC#1#2{\leavevmode\kern-.em
\raise.5ex\hbox{\the\scriptfont0 #1}\kern-.em
/\kern-.15em\lower.25ex\hbox{\the\scriptfont0 #2}}
\newif\ifrefphysrev

\def\refNP{\refphysrevfalse
           \typeout{** Reference: Nucl Phys format}}
\refphysrevtrue

\def \vol(#1,#2,#3){\ifrefphysrev{{\bf {#1}}, 
{#3} (19{#2})}\else{{{\bf {#1}}(19{#2}){#3}}}\fi}
\def \NP(#1,#2,#3){Nucl.\ Phys.\          \vol(#1,#2,#3)}
\def \PL(#1,#2,#3){Phys.\ Lett.\          \vol(#1,#2,#3)}
\def \PRL(#1,#2,#3){Phys.\ Rev.\ Lett.\   \vol(#1,#2,#3)}
\def \PRp(#1,#2,#3){Phys.\ Rep.\          \vol(#1,#2,#3)}
\def \PR(#1,#2,#3){Phys.\ Rev.\           \vol(#1,#2,#3)}
\def \PTP(#1,#2,#3){Prog.\ Theor.\ Phys.\ \vol(#1,#2,#3)}
\def \ibid(#1,#2,#3){{\it ibid.}\         \vol(#1,#2,#3)}
\makeatletter
\def\scriptsize{\@setsize\scriptsize{14.5pt}\xipt\@xipt
\abovedisplayskip 11\p@ plus3\p@ minus6\p@
\belowdisplayskip \abovedisplayskip
\abovedisplayshortskip  \z@ plus3\p@
\belowdisplayshortskip  6.5\p@ plus3.5\p@ minus3\p@
\def\@listi{\leftmargin\leftmargini
\parsep 4.5\p@ plus2\p@ minus\p@ \itemsep \parsep
\topsep 9\p@ plus3\p@ minus5\p@}}
\def \@magscale#1{ scaled \magstep #1}
\def \half(#1){\mathbox(\frac{#1}{2})}
\def \ninej(#1,#2,#3,#4,#5,#6,#7,#8,#9){\mathbox(\left\{\matrix 
     {#1&#2&#3\cr#4&#5&#6\cr#7&#8&#9\cr}\right\})}
\newif\ifnoncomplete

\noncompletefalse

\def\@cite#1#2{\unskip\nobreak\relax
    {[#1]}} 

\def\citenum#1{{\def\@cite##1##2{##1}\cite{#1}}}
\def\citea#1{\@cite{#1}{}}
 
\newcount\@tempcntc
\def\@citex[#1]#2{\if@filesw\immediate\write\@auxout{%
\string\citation{#2}}\fi
  \@tempcnta\z@\@tempcntb\m@ne\def\@citea{}\@cite{\@for\@citeb:=#2\do
    {\@ifundefined
       {b@\@citeb}{\@citeo\@tempcntb\m@ne\@citea\def\@citea{,}%
{\bf ?}\@warning
       {Citation `\@citeb' on page \thepage \space undefined}}%
{\setbox\z@\hbox{\global\@tempcntc0\csname b@\@citeb\endcsname\relax}%
     \ifnum\@tempcntc=\z@ \@citeo\@tempcntb\m@ne
       \@citea\def\@citea{,}\hbox{\csname b@\@citeb\endcsname}%
     \else
      \advance\@tempcntb\@ne
      \ifnum\@tempcntb=\@tempcntc
      \else\advance\@tempcntb\m@ne\@citeo
      \@tempcnta\@tempcntc\@tempcntb\@tempcntc\fi\fi}}\@citeo}{#1}}
\def\@citeo{\ifnum\@tempcnta>\@tempcntb\else\@citea\def\@citea{,}%
  \ifnum\@tempcnta=\@tempcntb\the\@tempcnta\else
   {\advance\@tempcnta\@ne\ifnum\@tempcnta=\@tempcntb %
\else \def\@citea{--}\fi
    \advance\@tempcnta\m@ne\the\@tempcnta\@citea\the\@tempcntb}\fi\fi}

\newcommand{\AmS}{{\protect\the\textfont2
  A\kern-.1667em\lower.5ex\hbox{M}\kern-.125emS}}

\title{Effects of Instantons on the YN Interaction}

\author{Sachiko~Takeuchi\address{
Japan College of Social Work, Kiyose, Tokyo 204-8555, Japan}
\thanks{Email: sachi-ta@sophia.ac.jp}, 
Yoshihiro~Tani\address{Department of Physics,
        Tokyo Institute of Technology,\\
 Meguro, Tokyo 152-8551, Japan},
and Makoto~Oka$^{\rm b}$
}

 \makeatother
\refNP
   \input epsf           
\begin{document}
\maketitle

\begin{abstract}
We investigate the symmetric and anti-symmetric spin-orbit forces (SLS and ALS)
of the effective $\Lambda$N interaction 
derived from a quark cluster model 
with the instanton-induced interaction (\III),
which can reproduce the
 observed YN cross sections as well as the observed NN scattering data.

It is found that coupling to the $\Sigma$N 
channel enhances $\Lambda$N ALS,
and therefore 
that the cancellation between SLS and ALS in the $\Lambda$N channel
becomes more complete.
This may be one of the major reasons why the
single-particle spin-orbit force 
of $\Lambda$ in nuclei is weak.
%
\end{abstract}

\section{Introduction}

 A valence quark model with \III\ 
contains four terms in the quark hamiltonian: 
the kinetic term, the confinement term, and the one-gluon exchange (OGE) term,
and the \III\ term.
It was found that this model can reproduce 
rough feature of the YN systems as well as the NN scattering data 
\cite{PTPals}.
In ref.\ \cite{Ta94}, we demonstrated that 
a valence quark model with \III\ gives the small LS splitting of the excited baryons
and the large LS force between two-nucleons,
which are difficult to be described simultaneously.
Thus,
introducing \III\ enables us to see 
the spin-orbit force from a more fundamental viewpoint.
In this work, we discuss the spin-orbit problem in 
the YN system.
%
%

\section{Model}

Here we employ a valence quark model 
to investigate the properties of the spin-orbit force in the $\Lambda$N systems.
The short range feature is supplied by the quark degrees of freedom 
with the
gluonic potentials.
The intermediate and long-range interaction is supplied by the meson-exchange: the
flavor-singlet and octet scalar mesons, the pseudo-scalar, and the vector mesons.
We use some of the coupling constants and masses of the mesons as fitting parameters,
which are taken to reproduce
the NN phase shifts and the low-energy $\Lambda$p cross section
and the phase shifts. 
The NN phase shift can be reproduced well: all the partial wave 
phase shifts at $E_{\rm cm}=5 \sim 150$ MeV 
do not deviate more than 4 degrees 
away from the results of the phase-shift analysis  up to the $F$-wave.
The $\Lambda$p $^1S_0$ phase shift is about 10 degrees higher than that of $^3S_1$,
which is suggested by the experiments \cite{PTPals}.

\section{$\Lambda$N spin-orbit force}

In a quark-meson hybrid model,
there are two origins
of ALS.
One comes from the gluonic interaction;
the {\it symmetric} LS force between quarks produces both of
the {\it antisymmetric} and the symmetric LS between baryons.
The other is the meson-exchange interaction;
the tensor couplings of the vector-meson exchange 
can produce ALS \cite{Ok98}.
Both of them 
survive at the flavor SU(3) limit and therefore are rather large.

From the observed levels of $\Lambda$-hypernuclei, it is believed
that the $\Lambda$N spin-orbit force is small comparing to that between
two nucleons \cite{RSY99}. 
Since the spin of u- and d-quark in the $\Lambda$ particle
is zero,
only the s-quark contributes to the $\Lambda$N spin-orbit force.
Because of this, 
the $\Lambda$N spin-orbit force (= $\Lambda$N SLS + $\Lambda$N ALS) is
small in a quark model, though each SLS or ALS can be large.
As seen in table 1, however, this cancellation
seems not enough.
Moreover, strong ALS seen in the model can cause
a bump in the $\Lambda$p elastic cross section, which is excluded
by the experiments.

\begin{figure}
\begin{minipage}{8cm}
\includegraphics*[scale=1]{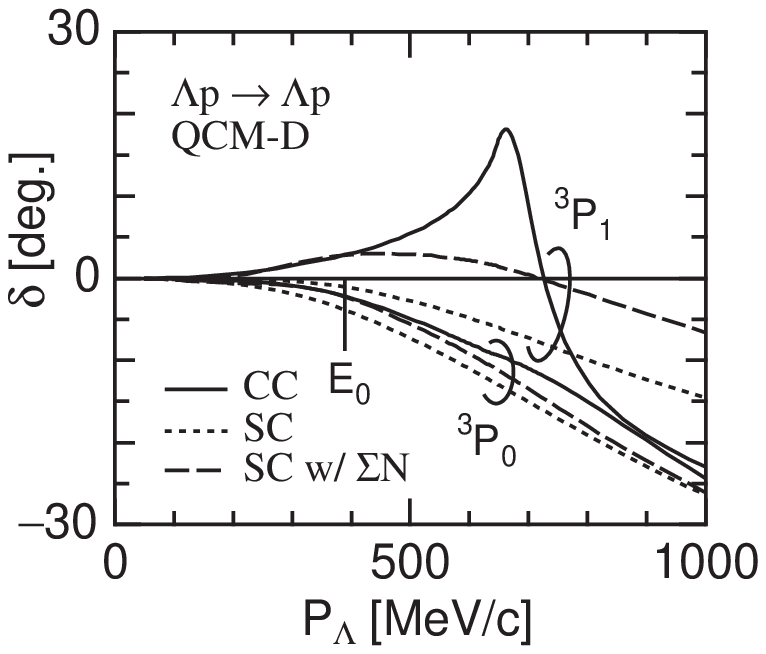}
\end{minipage}
\begin{minipage}{7.7cm}
\rule{0cm}{2cm}\\
\noindent
Figure 1. Phase shifts of the $\Lambda$N $^{3}P_{0}$ and $^3P_1$ channels.\\
The solid lines are for the $\Lambda$N-$\Sigma$N coupled-channel 
calculation, the dotted lines for the $\Lambda$N single-channel 
calculation, the dashed lines for the single-channel calculation 
by $\tilde H_{\Lambda\rmN}$
 with $E_{0}$ = 10 MeV 
(see text).
\end{minipage}
\vspace*{-0.5cm}
\end{figure}

\begin{table}[b]
Table 1. LS matrix elements by the quark cluster wave 
        functions with $b_{\rm baryon}$=1.35 fm.
\\
\begin{tabular}{l|cc|c|cc|c}
	\hline
	\hline
\rule{0cm}{5mm}Models 
& $H_{SLS}$ & $H_{ALS}$ & $H_{LS}$ 
& $\delta\tilde{H}_{SLS}$ & $\delta\tilde{H}_{ALS}$ 
& $\tilde{H}_{LS}$ \\
	\hline
QCM-B& $-$1.03 & 0.31 & $-$0.72 & 0.58 & 0.67 & 0.53\\
QCM-C& $-$1.22 & 0.26 & $-$0.95 & 0.56 & 0.50 & 0.10\\
QCM-D& $-$1.20 & 0.22 & $-$0.98 & 0.54 & 0.45 & 0.01\\
	\hline
	\hline
\end{tabular}
\\[0pt]
All entries are in MeV.
\end{table}

\section{Coupling to the $\Sigma$N channel}

The RGM equation of the quark cluster model,
$
	(H-EN) \psi  =  0 
$,
can be 
rewritten as the ``Schr\"odinger'' equation,
$
	(\overline H-E) \psibar = 0 
$,
with
$
\overline H  =  N^{-1/2}H N^{-1/2}$ and 
$	\psibar  =  N^{1/2} \psi
$.
The coupling to the $\Sigma$N channel is large in the quark model.
To  $\Lambda$N LS force should be modified as:
\begin{eqnarray}
\tilde H_{\Lambda\rmN} &=& \overline H_{11} 
- \overline H_{12} 
(\overline H_{22} - E_{0})^{-1}
\overline H_{21} 
	\label{LNA} 
\end{eqnarray}
where the channel 1 [2] denotes the $\Lambda$N [$\Sigma$N] channel.
At $E_{0}=E$, eq.\ (\ref{LNA}) is equivalent to the coupled-channel 
calculation.
As $E_{0}$ deviates from $E$, the result deviates from the exact one, 
which can be seen, {\it e.g.}\ in the scattering phase 
shift (Fig.\ 1).

\begin{figure}
\begin{minipage}{16cm}
\begin{center}
	\includegraphics*[scale=0.75]{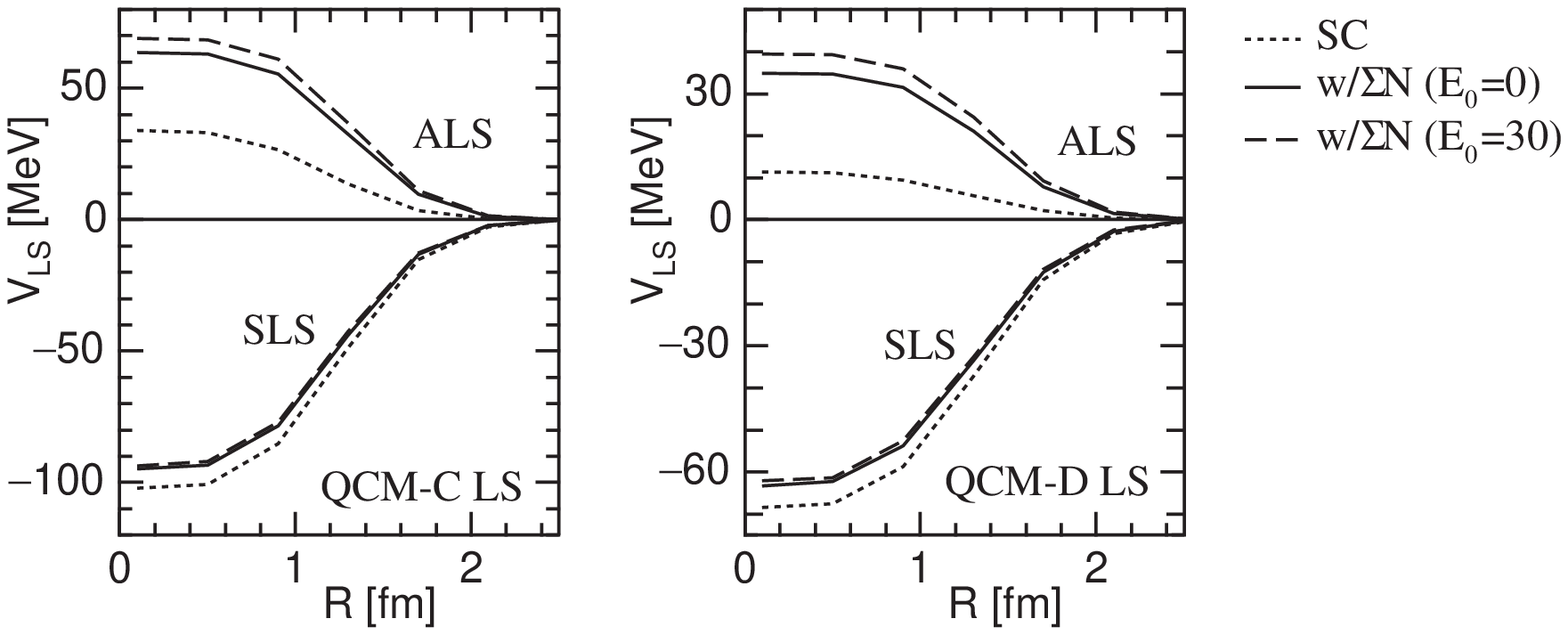}
\end{center}
\vspace*{-0.2cm}
\noindent
Figure 2. Adiabatic LS potentials for the $\Lambda$N $P$-wave channels.\\
The solid and the dashed lines are 
the single-channel calculation 
by $\tilde H_{\Lambda\rmN}$ with $E_{0}$ = 0 and 30 MeV
and the dotted lines for the $\Lambda$N single-channel calculation.
\vspace*{-0.5cm}
\end{minipage}
\end{figure}

To see rough size of the effect of the coupling to the $\Sigma$N channel,
we evaluate the hamiltonian $\tilde H_{\Lambda\rmN}$
by the quark-cluster wave functions. Their size 
corresponds to the ones with $b_{\rm baryon}$=1.35 fm, with an extra potential
between the baryons.
In table 1, we list that of SLS and ALS separately 
by three parameter sets of QCM \cite{PTPals}
(QCM-B has the meson ALS similar to NSC97f,
QCM-C [D] without [with] \III\ with no meson ALS).
$H_{SLS}$ and $H_{ALS}$ stand for the SLS and ALS terms in the 
$\Lambda$N force,
whereas $\delta \tilde H_{\alpha}$ stands for the
contribution from the $\Sigma$N channel to each LS term.
The size of the total LS, which consists of SLS and ALS with 
the $\Sigma$N effect, is listed under the entry $\tilde H_{\alpha}$.

The effect of the $\Sigma$N channel
is important;
it reduces SLS ($\delta\tilde{H}_{SLS}$) and 
enhances ALS ($\delta\tilde{H}_{ALS}$) largely, 
so that the size of the whole LS term decreases 
considerably ($\tilde{H}_{LS}$).
The adiabatic LS potentials obtained from the eq.\ (\ref{LNA})
are shown in Fig.\ 2, which also indicate 
that the $\Sigma$N channel effect is large.

\section{Summary}

In this work, we investigate the effect of coupling to the $\Sigma$N 
channel on the spin-orbit force in the $\Lambda$N interaction.
The effect seems important, which reduces the $\Lambda$ 
single particle LS force largely.


\end{document}